\documentclass[runningheads]{llncs}
\usepackage{graphicx}
\makeatletter
\RequirePackage[bookmarks,unicode,colorlinks=true]{hyperref}%
   \def\@citecolor{blue}%
   \def\@urlcolor{blue}%
   \def\@linkcolor{blue}%

\def\orcidID#1{\smash{\href{http://orcid.org/#1}{\protect\raisebox{-1.25pt}{\protect\includegraphics{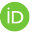}}}}}
\makeatother

\graphicspath{{figures/}{./figures/}}

\usepackage{marvosym}

\usepackage{mathtools, amsmath, amssymb, colonequals}
\usepackage[algoruled, linesnumbered, longend, noline]{algorithm2e}

\usepackage{wrapfig}
\usepackage{microtype}
\usepackage{mdframed}
\usepackage{hhline}
\usepackage{color,soul}
\usepackage{multirow}
\usepackage{paralist}

\usepackage[firstpage]{draftwatermark}
\SetWatermarkText{\hspace*{4.2in}\raisebox{6.3in}{\includegraphics[scale=0.1]{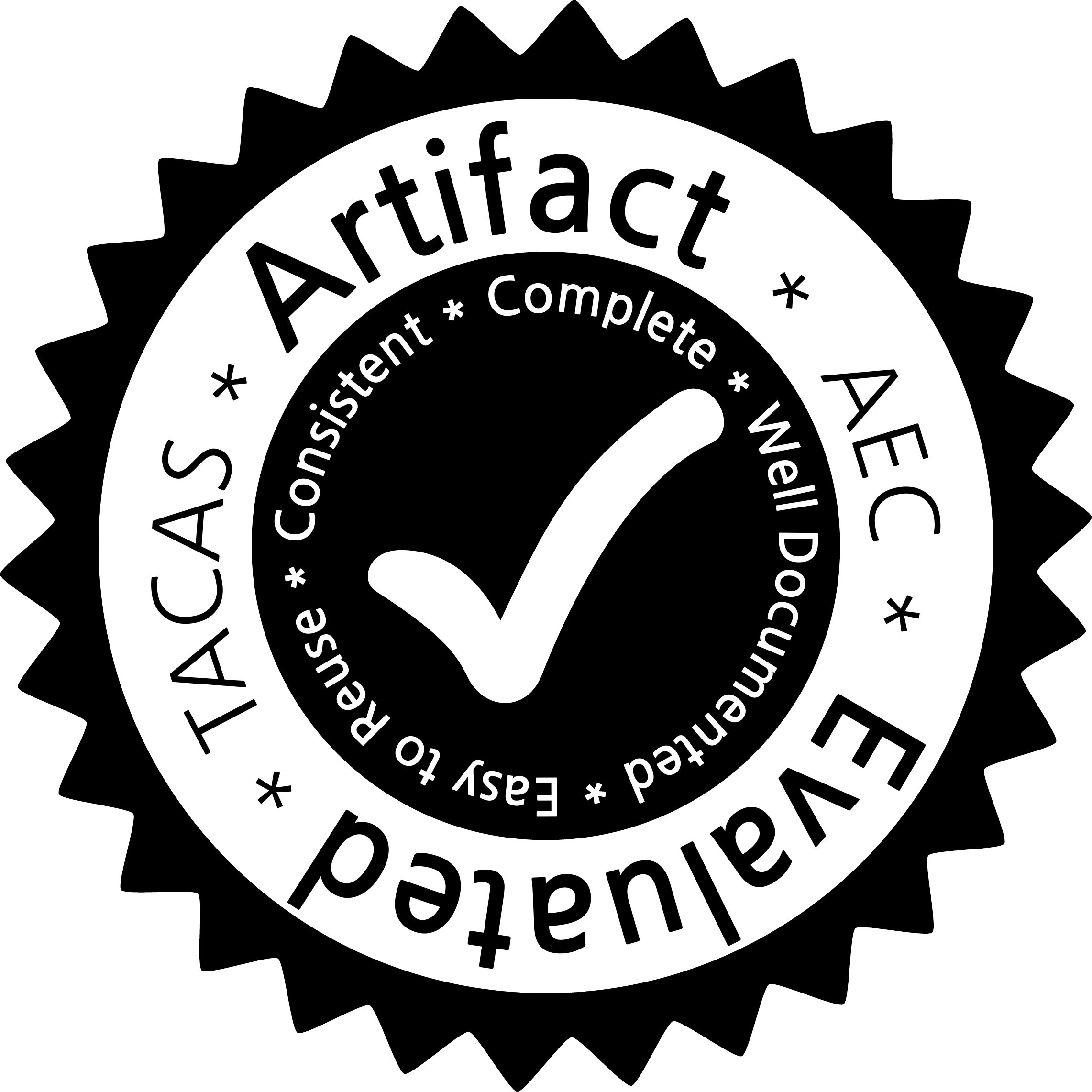}}}
\SetWatermarkAngle{0}

\usepackage[dvipsnames]{xcolor}

\usepackage{tikz}
\usetikzlibrary{automata,positioning,shapes,fit,calc}

\newcommand{\unitinterval}{[0,1]}

\DeclarePairedDelimiter\abs{\lvert}{\rvert}

\DeclareMathOperator*{\argmax}{arg\,max}

\newcommand{\vect}[1]{\boldsymbol{#1}}

\newcommand{\prob}{\mathbb{P}}
\newcommand{\supp}{\mathrm{supp}}

\renewcommand{\phi}{\varphi}
\newcommand{\F}[1]{\Diamond#1}
\newcommand{\reachability}[3]{\prob_{#1 #2}[\F{#3}]}
\newcommand{\safety}[2]{\reachability{\leq}{#1}{#2}}

\newcommand{\safetys}{\safety{\lambda}{T}}

\newcommand{\sins}{s \in S}
\newcommand{\sinit}{s_{0}}
\newcommand{\sBot}{s_{\bot}}
\newcommand{\sTop}{s_{\top}}

\renewcommand{\pm}{\vect{P}}
\newcommand{\mc}{(S,\sinit,\pm)}
\newcommand{\subchain}[2]{#1 {\downarrow} #2}

\newcommand{\vz}{\vect{0}}
\newcommand{\vo}{\vect{1}}

\newcommand{\vg}{\vect{\gamma}}

\newcommand{\fml}{\mathcal{D}}
\newcommand{\fpm}{\mathcal{B}}
\newcommand{\family}{(S,\sinit,K,\fpm)}

\newcommand{\rlz}{\mathcal{R}}
\newcommand{\rlzf}{\rlz^{\fml}}
\newcommand{\fmlr}{\fml_r}

\newcommand{\generalize}[2]{#1 {\uparrow} #2}
\newcommand{\generalizes}{\generalize{r}{K_C}}

\newcommand{\successRate}[1]{\sigma_{#1}}

\newcommand{\hermantwo}{Herman$^*$}

\begin{document}
\title{
Inductive Synthesis for Probabilistic Programs Reaches New Horizons\vspace{-0.5em}\thanks{This work has been partially supported by the Czech Science Foundation grant \mbox{GJ20-02328Y} and the ERC AdG Grant 787914 FRAPPANT, the NSF grants 1545126 (VeHICaL) and 1646208, by the DARPA Assured Autonomy program, by Berkeley Deep Drive, and by Toyota under the iCyPhy center.}
}
%
%

\author{Roman Andriushchenko\inst{1}\orcidID{0000-0002-1286-934X} \and
Milan \v{C}e\v{s}ka  (\Letter) \inst{1}\orcidID{0000-0002-0300-9727} \and\\
Sebastian Junges\inst{2}\orcidID{0000-0003-0978-8466} \and Joost-Pieter Katoen\inst{3}\orcidID{0000-0002-6143-1926}}

\authorrunning{R. Andriushchenko et al.}
%
\institute{
Brno University of Technology, Brno, Czech Republic \\
\email{ceskam@fit.vutbr.cz}
\and
University of California, Berkeley, USA
\and
RWTH Aachen University, Aachen, Germany
}
\maketitle    
\setcounter{footnote}{0}
\vspace{-1em}
\begin{abstract}
This paper presents a novel method for the automated synthesis of probabilistic programs. The starting point is a program sketch representing a finite family of finite-state Markov chains with related but distinct topologies, and a reachability specification. The method builds on a novel inductive oracle that greedily generates counter-examples (CEs) for violating programs and uses them to prune the family. These CEs leverage the semantics of the family in the form of bounds on its best- and worst-case behaviour provided by a deductive oracle using an MDP abstraction. The method further monitors the performance of the synthesis and adaptively switches between inductive and deductive reasoning. Our experiments demonstrate that the novel CE construction provides a significantly faster and more effective pruning strategy leading to an accelerated synthesis process on a wide range of benchmarks. For challenging problems, such as the synthesis of decentralized partially-observable controllers, we reduce the run-time from a day to~minutes.  
\end{abstract}

\section{Introduction}

\label{sec:introduction}
\paragraph{Background and motivation.}
Controller synthesis for Markov decision processes (MDPs~\cite{Put94}) and temporal logic constraints is a well-understood and tractable problem, with a plethora of mature tools providing efficient solving capabilities.
However, the applicability of these controllers to a variety of systems is limited: Systems may be decentralized, controllers may not be able to observe the complete system state, cost constraints may apply, and so forth. 
Adequate operational models for these systems exist in the form of \emph{decentralized partially-observable MDPs} (DEC-POMDPs~\cite{DBLP:series/sbis/OliehoekA16}). 
The controller synthesis problem for these models is undecidable~\cite{DBLP:conf/aaai/MadaniHC99}, and tool support (for verification tasks) is scarce. 

This paper takes a different approach: the controller together with the environment can be modelled as probabilistic program sketches where ``holes'' in the probabilistic program model choices that the controller may make. 
Conceptually, the controllers of the DEC-POMDP are described by a user-defined finite family~$\mathcal{M}$ of Markov chains. 
\emph{The synthesis problem that we consider is to find a Markov chain $M$ (i.e., a probabilistic program) in the family $\mathcal{M}$, such that $M \models \varphi$, where $\varphi$ is the specification.}
To allow efficient algorithms, the family must have some structure. 
In particular, in our setting, the family is parameterized by a set of discrete \emph{parameters} $K$; an assignment $K \rightarrow V$ of these parameters with concrete values $V$ from its associated domain yields a family member, i.e., a Markov chain (MC).
Such a parameterization is naturally obtained from the probabilistic program sketch, where some constants (or program parts) can be left open.  
The search for a family member can thus be considered as the search for a hole-assignment. 
This approach fits within the realm of syntax-guided synthesis~\cite{sygus}.

\medskip\noindent\emph{Motivating example.}
\emph{Herman's protocol}~\cite{herman-1} is a well-studied randomized distributed algorithm aimed to obtain fast stabilization on average.
In \cite{herman-2}, a family~$\mathcal{M}$ of MCs is used to model different protocol instances.
They considered each instance separately, and found which of the controllers for Herman's protocol performs best.
Let us consider the protocol in a bit more detail: It considers self-stabilization of a unidirectional ring of network stations where all stations have to behave similarly---an anonymous network. 
Each station stores a single bit, and can read the internal bit of one (say left) neighbour.
To achieve stabilization, a station for which the two legible bits coincide updates its own bit based on the outcome of a coin flip.
The challenge is to select a controller that flips this coin with an optimal bias, i.e., minimizing the expected time until stabilization. 
In a setting where the probabilities range over $0.1, 0.2, \ldots, 0.9$, this results in analyzing nine different MCs.
Does the expected time until stabilization reduce if the controllers are additionally allowed to have a single bit of memory? 
In every step, there are $9{\cdot}9$ combinations for selecting the coin flip and for each memory cell and coin flip outcome, the memory can now be updated, yielding $2{\cdot}2{\cdot}2$ possibilities. 
This one-bit extension thus results in a family of $648$ models.
If, in addition, one allows stations to make decisions depending on the token-bits, both the coin flips and the memory updates are multiplied by a factor $4$, yielding $10,368$ models. 
Eventually, analyzing all individual MCs is infeasible. 

\medskip\noindent\emph{Oracle-guided synthesis.}
To tackle the synthesis problem, we introduce an \emph{oracle-guided inductive synthesis} approach~\cite{10.1145/1806799.1806833,Solar-LezamaPLDI2005}. 
A learner selects a family member and passes it to the oracle.
The oracle answers whether the family member satisfies $\varphi$, and crucially, gives additional information in case this is not the case. 
Inspired by~\cite{cegis}, if the family member violates the specification $\varphi$, our oracle returns a set $K'$ of parameters such that all family members obtained by changing only the values assigned to $K'$ violate $\varphi$. 
We argue that such an oracle must (1) induce little overhead in providing $K'$, (2) be aware of the existence of parameters in the family, and (3) have (resemblance of) awareness about the semantics of the parameters and their values.

\medskip\noindent\emph{Oracles.}
With these requirements in mind, we construct a counterexample (CE)-based oracle from scratch. 
We do so by carefully exploiting existing methods.
We construct critical subsystems as CEs~\cite{DBLP:conf/sfm/AbrahamBDJKW14}. 
Critical subsystems are parts of the MC that suffice to refute the specification. 
If a hole is absent in a CE, its value is irrelevant. 
To avoid the cost of finding optimal CEs---an NP-hard problem~\cite{DBLP:conf/tacas/FunkeJB20}---we consider greedy CEs that are similar to~\cite{cegis}. 
However, our greedy CEs are aware of the parameters, and try to limit the occurrence of parameters in the CE. 
Finally, to provide awareness of the semantics of parameter values, we provide lower and upper bounds on all states: Their difference indicates how much varying the value at a hole may change the overall reachability probability.
These bounds are efficiently computed by another oracle. 
This oracle analyses a quotient MDP obtained by employing an abstraction method that is part of the abstraction-refinement loop in~\cite{cegar}.

\medskip\noindent\emph{A hybrid variant.}
The two oracles are significantly different. 
Abstraction refinement is \emph{deductive}: it argues about single family members by considering (an aggregation of) all family members.
The critical subsystem oracle is \emph{inductive}: by examining a single family member, it infers statements about other family members.
This suggests a middle ground: a \emph{hybrid strategy} monitors the performance of the two oracles during the synthesis and suggests their best usage. 
More precisely, the hybrid strategy integrates the counterexample-based oracle into the abstraction-refinement loop. 

\medskip\noindent\emph{Major results.}
We present a novel and dedicated oracle deployed in an efficacious synthesis loop. 
We use model-checking results on an abstraction to tailor smaller CEs.
Our greedy and family-aware CE construction is substantially faster than the use of optimal CEs. 
Together, these two improvements yield CEs that are on par with optimal CEs, but are found much faster.
The integration of multiple abstraction-refinement steps yields a superior performance:x
We compare our performance with the abstraction-refinement loop from~\cite{cegar} using benchmarks from~\cite{cegar}. 
Benchmarks can be classified along two dimensions: 
($A$) Benchmarks with a structure good for CE-generation.
($B$) Benchmarks with a structure good for abstraction-refinement.
A-benchmarks are a natural strength of our novel oracle.
Our simple, efficient hybrid strategy significantly outperforms the state-of-the-art on $A$-benchmarks, while it only yields limited overhead for $B$-benchmarks. 
Most importantly, the novel hybrid strategy can solve benchmarks that are out of reach for pure abstraction-refinement or pure CE-based reasoning. 
In particular, our hybrid method is able to synthesize the optimal Herman protocol with memory---the synthesis time on a design space with 3.1~millions of candidate programs reduces from a day to minutes.

\subsubsection{Related work}
The synthesis problems for parametric probabilistic systems can be divided into the following two categories. 

\medskip\noindent\emph{Topology synthesis,} akin to the problem considered in this paper, assumes a finite set of parameters affecting the MC topology. Finding an instantiation satisfying a reachability property is NP-complete in the number of parameters~\cite{10.1007/978-3-030-30806-3_7}, and can naively be solved by analyzing all individual family members.  An alternative is to model the MC family by an MDP and resort to standard MDP model-checking algorithms. Tools such as ProFeat~\cite{DBLP:journals/fac/ChrszonDKB18} or QFLan~\cite{DBLP:conf/fm/VandinBLL18} take this approach to quantitatively analyze alternative designs of software product lines~\cite{DBLP:journals/infsof/GhezziS13,lanna2018feature}. These methods are limited to small families. This motivated 
(1) \emph{abstraction-refinement} over the MDP representation~\cite{cegar}, and (2) \emph{counterexample-guided inductive synthesis} (CEGIS) for MCs~\cite{cegis}, mentioned earlier. The alternative problem of sketching for probabilistic programs that fit given data is studied, e.g., in~\cite{NoriPLDI2015,saad2019bayesian}.

\medskip\noindent\emph{Parameter synthesis} considers models with uncertain parameters associated to transition probabilities, and analyses how the system behaviour depends on the~parameter values. The most promising techniques are based on \emph{parameter lifting} that treats identical parameters in different transitions independently~\cite{acta,Quatmann2016} and has been implemented in the state-of-the-art probabilistic model checkers Storm~\cite{STORM} and PRISM~\cite{KNP11}. 
An alternative approach based on building rational
functions for the satisfaction probability has been proposed in~\cite{DBLP:conf/ictac/Daws04} and further improved in~\cite{hahn2011probabilistic,dehnert2015prophesy,DBLP:journals/iandc/BaierHHJKK20}. This approach has been also applied to different problems such as model repair~\cite{model-repair-1,pathak-et-al-nfm-2015,DBLP:journals/iandc/Chatzieleftheriou18}. 

Both synthesis problems can be also attacked by \emph{search-based techniques} that do not ensure an exhaustive exploration of the parameter space. 
These include evolutionary techniques~\cite{Harman2012,Martens:2010:AIS:1712605.1712624} and genetic algorithms~\cite{DBLP:journals/ase/GerasimouCT18}.
Combinations with parameter synthesis have been used~\cite{CALINESCU2018140} to synthesize robust systems.

\section{Problem Statement}
\label{sec:preliminaries}

We formalize the essential ingredients and the problem statement. See \cite{Baier2018} for more material.

\paragraph{Sets of Markov chains.}
A \emph{(discrete) distribution} over a finite set $X$ is a~function $\mu \colon S \rightarrow \unitinterval$~s.t.~$\sum_x \mu(x) = 1$. The set $Distr(X)$ contains all distributions over $X$. The~\emph{support} of $\mu \in Distr(X)$ is $\supp(\mu) = \{ x \in X \mid \mu(x) > 0\}$.
\begin{definition}[MC]
\label{def:dtmc}
A \emph{Markov chain (MC)} is a~tuple $D = \mc$, where $S$ is a~finite set of \emph{states}, $\sinit \in S$ is an~\emph{initial state}, and $\pm\colon S \rightarrow Distr(S)$ is a~\emph{transition probability function}. We write $\pm(s,t)$ to denote $\pm(s)(t)$. The~state~$s$ is \emph{absorbing} if $\pm(s,s) = 1$.
\end{definition}

Let $K$ denote a finite set of discrete parameters with finite domain $V_k$. 
For brevity, we often assume that all domains are the same, and omit the subscript $k$. 
A \emph{realization} $r$ maps parameters to values in their domain, i.e., $r\colon K \rightarrow V$. 
Let $\rlzf$ denote the set of all realizations of a set  $\mathcal{D}$ of MCs.
A $K$-parameterized set of MCs $\mathcal{D}(K)$ contains the MCs $\mathcal{D}_r$, for every $ r \in \rlzf$.
In Sect.~\ref{sec:family}, we give an operational model for such sets. In particular, realizations will fix the targets of transitions. In our experiments, we describe these sets using the PRISM modelling language where parameters are described by undefined integer values.

\paragraph{Properties and specifications.}
 For simplicity, we consider (unbounded) \emph{reachability} properties\footnote{Our implementation also supports expected reachability rewards.}.
 For a~set~$T \subseteq S$ of \emph{target states}, let $\prob[D,s \models \F{T}]$ denote the~probability in MC $D$ to eventually reach some state in $T$ when starting in the~state~$\sins$. 
 A property $\phi \equiv \reachability{\bowtie}{\lambda}{T}$ with $\lambda \in [0,1]$ and $\bowtie \, \in \{ \leq, \geq\}$ expresses that the probability to reach $T$ does relate to $\lambda$ according to $\bowtie$. 
 If $\bowtie \, = {\leq}$, then $\phi$ is a \emph{safety} property; otherwise, it is a \emph{liveness} property. 
Formally, state $s$ in MC $D$ satisfies $\phi$ if ${\prob[D,s \models \F{T}] \geq \lambda}$. 
The~MC $D$ satisfies $\phi$ if the above holds for its~initial state.
A \emph{specification} is a set of properties $\Phi = \{\phi_i\}_{i \in I}$, and $D \models \Phi$ if $\forall i \in I: D \models \phi_i$. 

\paragraph{Problem statement.}
The key problem statement in this paper is \emph{feasibility}:
\begin{mdframed}[backgroundcolor=blue!10]
Given a parameterized set of Markov chains $\mathcal{D}(K)$ over parameters $K$ and a specification $\Phi$, find a realization $r\colon K \rightarrow V$ such that $\mathcal{D}_r \models \Phi$. 
\end{mdframed}

\noindent When $\mathcal{D}$ is clear from the context, we often write $r \models \Phi$ to denote $\mathcal{D}_r \models \Phi$.

We additionally consider the optimizing variant of the synthesis problem.
The \emph{maximal synthesis} problem asks: given a maximizing property $\phi_{\max} \equiv \reachability{\bowtie}{\lambda}{T}$, identify $r^* \in \argmax_{r \in \rlzf} \left \{ \prob[\fmlr \models \F{T}] \mid \fmlr \models \Phi \right \}$ provided it exists. 
The \emph{minimal synthesis} problem is defined analogously.

As the~state space $S$, the~set $K$ of parameters, and their domains are all finite, the above synthesis problems are decidable.
One possible solution, called the \emph{one-by-one approach}~\cite{onebyone}, considers each realization $r \in \rlzf$. 
The state-space and parameter-space explosion renders this approach unusable for large problems, necessitating the~usage of advanced techniques that exploit the family structure.

\section{Counterexample-Guided Inductive Synthesis}
In this section, we recap a baseline for a counterexample-guided inductive synthesis (CEGIS) loop, as put forward in~\cite{cegis}.
In particular, we first instantiate an oracle-guided synthesis method, discuss an operational model for families, giving structure to the parameterized set of Markov chains, and finally detail the usage of CEs to create an oracle. 

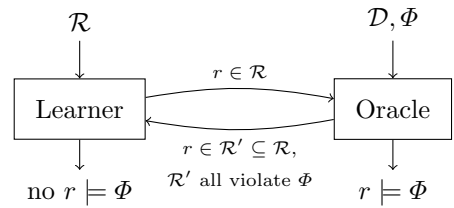
\begin{wrapfigure}[9]{r}{0.5\textwidth}
  \begin{center}
  \vspace{-3em}
    \begin{tikzpicture}
    \node[rectangle, draw, inner sep=8pt] (learner) {Learner};
    \node[rectangle, draw, inner sep=8pt,right=2.5cm of learner] (oracle) {Oracle};
    \node[above=0.5cm of learner] (rlz) {$\rlz$};
    \draw[->] (rlz) -- (learner);
    \node[above=0.5cm of oracle] (phi) {$\mathcal{D},\Phi$};
    \draw[->] (phi) -- (oracle);
    \draw[->] (learner) edge[bend left=10] node[above] {\scriptsize{$r \in \rlz$}} (oracle);
    \draw[->] (oracle) edge[bend left=10] node[below,align=center] {\scriptsize{$r\in \rlz' \subseteq \rlz$, }\\\scriptsize{$\rlz'$ all violate $\Phi$}} (learner);
    \node[below=0.4cm of oracle] (sat) {$r \models \Phi$};
    \draw[->] (oracle) -- (sat);
    \node[below=0.4cm of learner]   (unsat) {no $r \models \Phi$};
    \draw[->] (learner) -- (unsat);
    \end{tikzpicture}
  \end{center}
  \vspace{-1.5em}
   \caption{Oracle-guided synthesis}
    \label{fig:oracle}
\end{wrapfigure}

Consider Fig.~\ref{fig:oracle}. 
A learner takes a set $\rlz$ of realizations, and has to find a realization $\mathcal{D}_r$ satisfying the specification $\Phi$. 
The learner maintains (a symbolic representation of) a set $Q \subseteq \rlz$ of realizations that need to be checked.
It iteratively asks the oracle whether a particular $r \in Q$ is a solution. 
If it is a solution, the oracle reports success. 
Otherwise, the oracle returns a set $\rlz'$ containing $r$ and potentially more realizations all violating $\Phi$. 
The learner then prunes $\rlz'$ from $Q$.
In Section~\ref{sec:main}, we focus on creating an efficient oracle that computes a set $\rlz'$ (with $r \in \rlz'$) of realizations that are all violating $\Phi$. 
In Section~\ref{sec:adaptive}, we provide a more advanced framework that extends this method.
 The remainder of this section lays the groundwork for these sections.

\subsubsection{Families of Markov chains}
\label{sec:family}
To avoid the need to iterate over all realizations, an efficient oracle exploits some structure of the family. 
In this paper, we focus on sets of Markov chains having different topologies. 
We explain our concepts using the operational model of families given in~\cite{cegar}. 
Our implementation supports (more expressive) PRISM programs with undefined integer constants.

\begin{definition}[Family of MCs]
\label{def:family}
A \emph{family of MCs} is a~tuple $\fml = \family$ with $S$ and $\sinit$ as before, $K$ is a~finite set of parameters with domains $V_k  \subseteq S$ for each $k \in K$, and $\fpm : S \rightarrow Distr(K)$ is a~family of transition probability functions.
\end{definition}
Function $\fpm$ of a~family $\fml$ of MCs maps each state to a~distribution over parameters~$K$. In the~context of the~synthesis of probabilistic models, these parameters represent unknown options or features of a~system under design. Realizations are now defined as follows.

\begin{definition}[Realization]
\label{def:realization}
A \emph{realization} of a~family $\fml = \family$ of MCs is a~function $r: K \rightarrow S$ s.t.~$r(k) \in V_k$, for all $k \in K$. We say that realization~$r$ induces MC $\fmlr = (S,\sinit,\fpm_r)$ iff $\fpm_r(s,s') = \sum_{k \in K, r(k) = s'}\fpm(s)(k)$ for any pair of states $s,s' \in S$. The set of all realizations of $\fml$ is denoted as $\rlzf$.
\end{definition}
The~set $\rlzf = \prod_{k \in K} V_k$ of all possible realizations is exponential in $\abs{K}$.

\subsubsection{Counterexample-guided oracles}
\label{subsec:cegis}
We first consider the feasibility synthesis for a single-property specification and later, cf.\ Remark~\ref{remark:gen}, generalize this to multiple properties and to optimal synthesis.
The notion of counterexamples is at the heart of the oracle from~\cite{cegis} and Sect.~\ref{sec:main}.

If an MC $D \not\models \phi$, a \emph{counterexample} (CE) based on a critical subsystem can serve as diagnostic information about the source of the failure. 
We consider the following CE, motivated by the notion of critical subsystem in~\cite{ce-rewards}.
\begin{definition}[Counterexample]
\label{def:subchain}
Let $D = \mc$ be an~MC with $\sBot \not \in S$. The~\emph{sub-MC} of $D$ induced by $C \subseteq S$ is the~MC $\subchain{D}{C} = (S \cup \{\sBot\},\sinit,\pm')$, where the~transition probability function $\pm'$ is defined by:
\begin{align*}
\pm'(s) =
\begin{cases}
\pm(s) & \text{ if } s \in C, \\
[\sBot \mapsto 1] & \text{ otherwise}.
\end{cases}
\end{align*}
\\[-0.6em]
The set $C$ and the sub-MC $\subchain{D}{C}$ are called a \emph{counterexample} (CE) for the property $\safetys$ on MC $D$, if $\subchain{D}{C} \not \models \safety{\lambda}{(T \cap (C \cup \{\sinit\}))}$. 
\end{definition}
Let $\fmlr$ be an MC violating the specification $\phi$. 
To compute other realizations violating $\phi$, the oracle computes a critical subsystem $\subchain{\fmlr}{C}$, which is then used to deduce a so-called \emph{conflict} for $\fmlr$ and $\phi$.

\begin{definition}[Conflict]
\label{def:conflict}
For family of MCs $\fml = \family$ and $C \subseteq S$, the set $K_C$ of \emph{relevant parameters} (called \emph{conflict}) is given by $\bigcup_{s \in C}\supp(\fpm(s))$.
\end{definition}
It is straightforward to compute a set of violating realizations from a conflict.
A \emph{generalization} of realization $r$ induced by the set $K_C \subseteq K$ of relevant parameters is the set $\generalizes =  \{r' \in \rlz \mid \forall k \in K_C : $ $r(k) = r'(k)\} $.
We often use the term \emph{conflict} to refer to its generalization. 
The size of a conflict, i.e., the number $|K_C|$ of relevant parameters $K_C$ is crucial.
Small conflicts potentially lead to generalizing $r$ to larger subfamilies $\generalizes$. 
It is thus important that the CEs contain as few parameterized transitions as possible. 
The size of a CE in terms of the number of states is not of interest. 
Furthermore, the overhead of providing CEs should be bounded from below by the payoff: Finding a large generalization may take some time, but small generalizations should be returned quickly.
The CE-based oracle in~\cite{cegis} uses an off-the-shelf CE procedure~\cite{DBLP:conf/atva/DehnertJWAK14,DBLP:journals/corr/abs-1305-5055}, and mostly does not provide small CEs.

\section{A Smart Oracle with Counterexamples and Abstraction}
\label{sec:main}

This section develops an oracle based on CEs, tailored for the use in an oracle-guided inductive synthesis loop described in Sect.~\ref{sec:family}. 
Its main features are:
\begin{compactitem}
    \item a fast greedy approach to compute CEs that provide small conflicts: We achieve this by taking into account the position of the parameters.
    \item awareness about the semantics of parameters by using model-checking results from an abstraction of the family. 
\end{compactitem}
Before going into details, we provide some illustrative examples. 

\subsubsection{A motivating example}

\begin{figure}[t]%
\centering
\begin{minipage}{0.3\textwidth}
\includegraphics[width=1\textwidth]{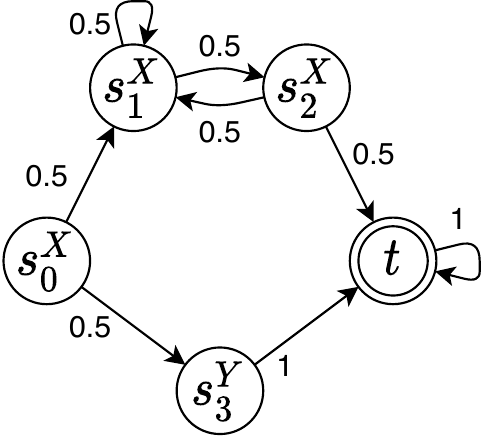}
\end{minipage}\hspace{1em}
\begin{minipage}{0.3\textwidth}
\includegraphics[width=1\textwidth]{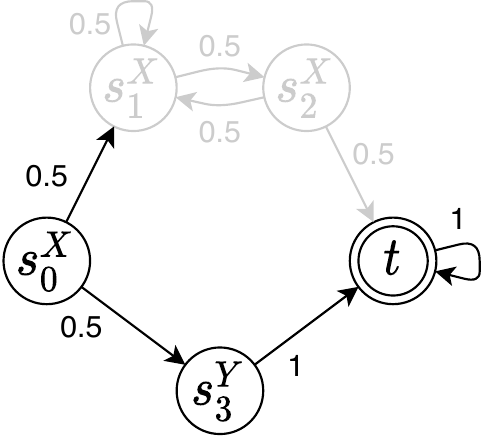}
\end{minipage}\hspace{1em}
\begin{minipage}{0.3\textwidth}
\includegraphics[width=1\textwidth]{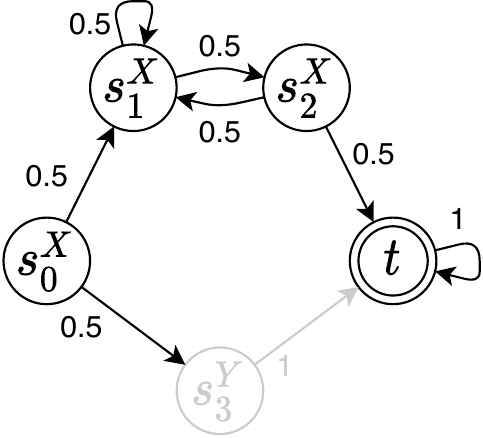}
\end{minipage}\hspace{1em}
\vspace{-1em}
\caption{Counterexamples for smaller conflicts.}
\label{fig:expansion}
\end{figure}
First, we illustrate what it means to take CEs that lead to small conflicts. 
Consider Fig.~\ref{fig:expansion}, with a family member $\fmlr$ (left), where the superscript of a state identifier $s_i$ denotes parameters relevant to $s_i$. 
Consider the safety property $\phi \equiv \safety{0.4}{\{t\}}$. 
Clearly, $\fmlr \not \models \phi$, and we can construct two CEs: $C_1 = \{s_0,s_3,t\}$ (center) and $C_2 = \{s_0,s_1,s_2,t\}$ (right) with conflicts $K_{C_1} = \{X,Y\}$ and $K_{C_2} = \{X\}$, respectively. 
It illustrates that a smaller CE does not necessarily induce a smaller conflict.

\begin{figure}[t]
\centering
\begin{minipage}{0.45\textwidth}
\begin{align*}
\fpm(\sinit) &= [X \mapsto 1], \\
\fpm(s_1) &= [T' \mapsto 0.6, Y \mapsto 0.2, F' \mapsto 0.2], \\
\fpm(s_2) &= [T' \mapsto 0.2, Y \mapsto 0.2, F' \mapsto 0.6], \\
\fpm(t) &= [T' \mapsto 1], \\
\fpm(f) &= [F' \mapsto 1]
\end{align*}
\end{minipage} \hspace{1em}
\begin{minipage}{0.5\textwidth}
\includegraphics[width=1\textwidth]{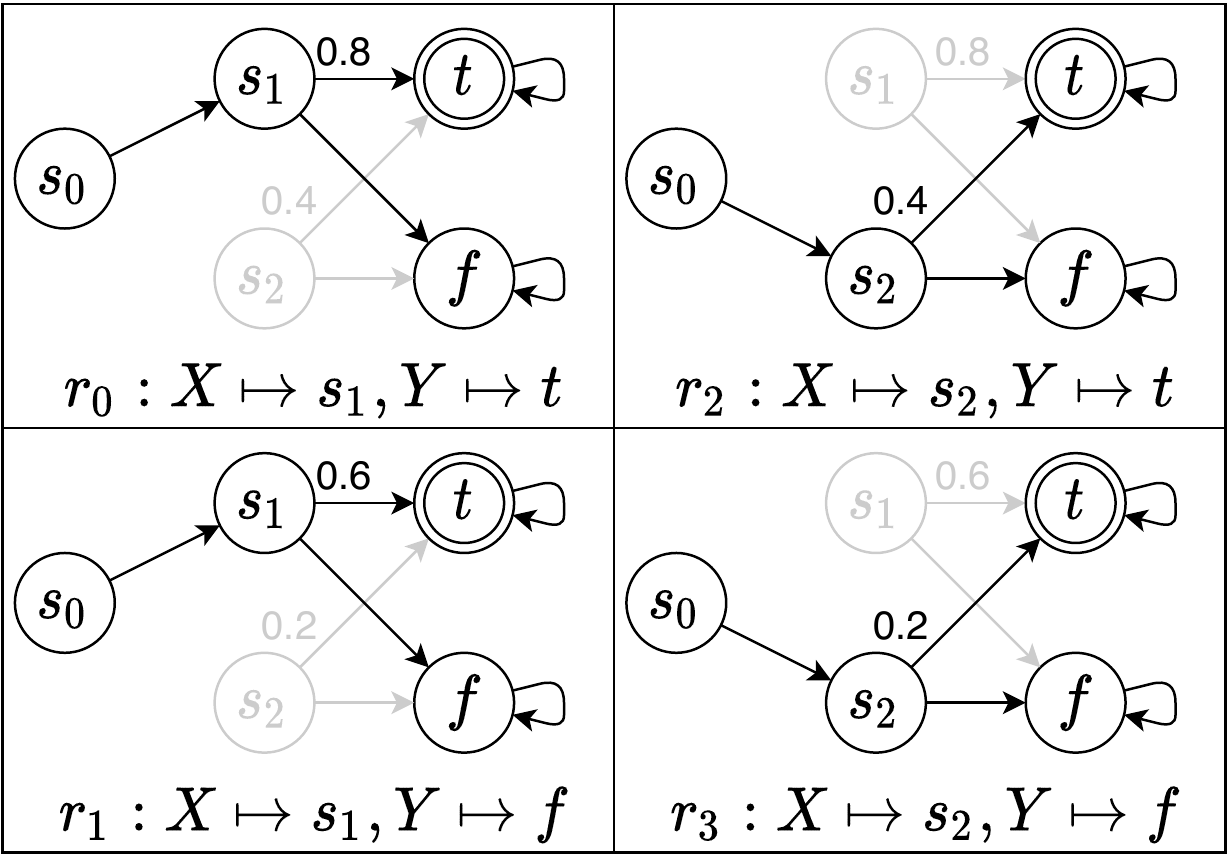}
\end{minipage}
\caption{A family $\fml$ of four Markov chains (unreachable states are grayed out).}
\label{fig:my_label}
\vspace{-1.5em}
\end{figure}

We now illustrate awareness of the semantics of parameters.
Consider the family $\fml = (S,\sinit,K',\fpm)$, where $S = \{\sinit,s_1,s_2,t,f\}$, the~parameters are $K' = \{X,Y,T',F'\}$ with domains $V_X = \{s_1,s_2\}$, $V_Y = \{t,f\}$, $V_{T'} = \{t\}$, $V_{F'} = \{f\}$, and a~family $\fpm$ of transition probability functions defined in Fig.~\ref{fig:my_label}~(left).
As the parameters $T'$ and $F'$ each can take only one value, we consider $K = \{X,Y\}$ as the~set of parameters. 
There are $\abs{V_X} \times \abs{V_Y} = 4$ family members, depicted in~Fig.~\ref{fig:my_label}(right). 
For conciseness, we omit some of the transition probabilities (recall that transition probabilities sum to one). 
Only realization~$r_3$ satisfies the safety property $\phi \equiv \safety{0.3}{\{t\}}$. 

\emph{CEGIS~\cite{cegis} illustrated}: Consider running CEGIS, and assume the oracle gets realization $r_0$ first. 
A model checker reveals $\prob[{\mathcal{D}_{r_0}},s_0 \models \F{T}] = 0.8 > 0.3$. 
The CE for $\fml_{r_0}$ and $\phi$ contains the (only) path to the target: $\sinit {\rightarrow} s_1 {\rightarrow} \, t$ having probability $0.8 > 0.3$. 
The~corresponding CE $C = \{\sinit,s_1,t\}$ induces the conflict $K_C = \{X,Y\}$. 
None of the~parameters is generalized. 
The~same argument applies to any subsequent realization: the~constructed CEs do not allow for generalization, the oracle returns only the passed realization, and the learner keeps iterating until accidentally guessing $r_3$.

\emph{Can we do better?} To answer this, consider CE generation as a game: The Pruner creates a critical subsystem $C$. 
The Adversary wins if it finds a MC satisfying $\phi$ containing $C$, thus refuting that $C$ is a counterexample.
In our setting, we change the game:
The Adversary must select a family member rather than an arbitrary MC. 
Analogously, off-the-shelf CE generators construct a critical subsystem $C$ that for every possible extension of $C$ is a CE. 
These are \emph{CEs without context}. 
In our game, the Adversary may not extend the MC arbitrarily, but must choose a family member.  
These are \emph{CEs modulo a family}.

\emph{Back to the example:} Observe that for a~CE for $\mathcal{D}_{r_0}$, we could omit states $t$ and $s_1$ from the set $C$ of critical states: we know for sure that, once $\mathcal{D}_{r_0}$ takes transition~$(\sinit,s_1)$, it will reach target state~$t$ with probability at least $0.6$.
This exceeds the threshold $0.3$, regardless of the~value of the~parameter~$Y$. 
Hence, for family $\fml$, the set~$C' = \{\sinit\}$ is a critical subsystem. 
The immediate advantage is that this set induces conflict $K_{C'} = \{X\}$ (parameter $Y$ has been generalized).
This enables us to reject all realizations from the set $\generalize{r_0}{K_{C'}} = \{r_0,r_1\}$. 
\emph{It is `easier' to construct a CE for a (sub)family than for arbitrary MCs}. 
More generally, a successful oracle needs to have access to useful bounds, and effectively integrate them in the CE generation.  

\label{subsec:ce}
\newcommand{\ub}{\textsl{ub}^\rlz}
\newcommand{\lb}{\textsl{lb}^\rlz}

\subsubsection{Counterexample construction}
We develop an algorithm using bounds on reachability probabilities, similar to the bounds used above. 
Let us assume that for some set of realizations $\rlz$ and for every state $s$, we have bounds $\lb(s), \ub(s)$, such that for every $r \in \rlz$ we have $\lb(s) \leq  \prob[\fmlr, s \models \F{T}] \leq \ub(s)$.
Such bounds always exist (take $0$ and $1$). We see later how we compute these bounds.
In what follows, we fix $r$ and denote $\mathcal{D}_r = \mc$.
Let us assume $\mathcal{D}_r$ violates a~safety property $\phi \equiv \reachability{\leq}{\lambda}{T}$. 
The following definition is central:
\begin{definition}[Rerouting]
\label{def:rerouting}
Let MC $D = \mc$ with $\sTop,\sBot \not \in S$, $C \subseteq S$ a~set of \emph{expanded states} and $\vg \colon S \setminus C \rightarrow \unitinterval$ a~\emph{rerouting vector}. 
The \emph{rerouting} of MC~$D$ w.r.t.~$C$ and~$\vg$ is the~MC $\subchain{D}{C}[\vg]  = (S \cup \{\sBot,\sTop\}, \sinit, \pm^{C}_{\vg})$ with:
\begin{align*}
\pm^{C}_{\vg}(s) =
\begin{cases}
\pm(s) & \text{ if } s\in C, \\
[\sTop \mapsto \vg(s),\sBot \mapsto (1{-}\vg(s))]  & \text{ if } s \in S {\setminus} C, \\
[s \mapsto 1] & \text{ if } s \in \{\sTop,\sBot\}. \\
\end{cases}
\end{align*}
\end{definition}
Essentially, $\subchain{D}{C}[\vg]$ extends the MC $D$ with additional \emph{sink states}~$\sTop$ and~$\sBot$ and replaces all outgoing transitions of any non-expanded state $s \in S {\setminus} C$ by a~transition leading to~$\sTop$ (with probability $\vg(s)$) and a complementary one to~$\sBot$. 
We consider $\sTop$ to be the new target and let $\varphi'$ denote the updated property. 
The transition $s\xrightarrow{\vg(s)} \sTop$ may be considered a `shortcut' that by-passes successors of $s$ and leads straight to target $\sTop$ with probability $\vg(s)$. 
To ensure that $\subchain{D}{C}[\vg]$ is a CE, the value $\vg(s)$ must be a lower bound on the reachability probability from $s$ in $D$. 
When constructing a CE for a singular MC, we pick $\vg = \vz$, whereas when this MC is induced by a realization $r \in \rlz$, we can safely pick $\vg = \lb$. 
The CE will be valid for every $r' \in \rlz$. 
It is a CE-modulo-$\rlz$.

Algorithmically, we employ a state-exploration approach and therefore start with $C^{(0)} = \emptyset$, i.e., all states are initially rerouted. 
If this is a CE, we are done. 
Otherwise, if the rerouting $\subchain{D}{C^{(0)}}[\vg]$ satisfies $\phi'$, then we `expand' some states to obtain a CE. 
Naturally, we must expand reachable states to change the satisfaction of $\phi$. 
By expanding some state $\sins$, we abandon the abstraction associated with the shortcut $s \xrightarrow{\vg(s)} \sTop$ and replace it with concrete behavior that was inherent to state $s$ in MC $D$. 
Expanding a state cannot decrease the induced reachability probability as $\lb$ is a valid lower bound. 
This gradual expansion of the reachable state space continues until for some $C \subseteq S$ the corresponding rerouting $\subchain{D}{C}[\vg]$ violates $\phi'$.
This gradual expansion process terminates as $\subchain{D}{S}[\vg] \equiv D$ and our assumption is $D \not \models \phi$.
We show this process on an example.

\begin{figure}[t]%
\centering
\begin{minipage}{0.35\textwidth}
\includegraphics[width=1\textwidth]{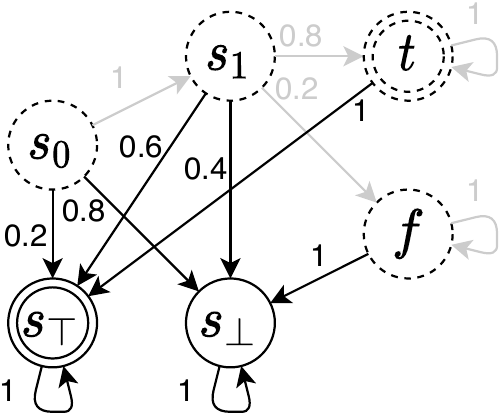}
\end{minipage}\hspace{1em}
\begin{minipage}{0.35\textwidth}
\includegraphics[width=1\textwidth]{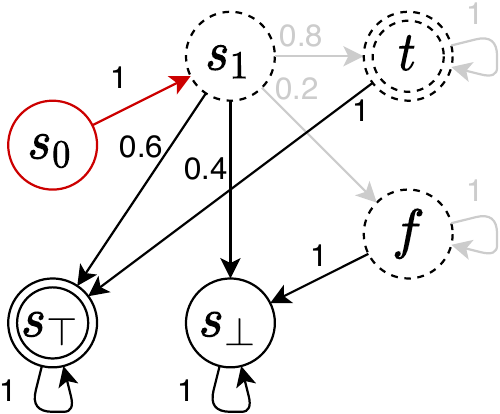}
\end{minipage}\hspace{1em}
\caption{Finding a~CE to $\fml_{r_0}$ and $\phi$ from Fig.~\ref{fig:my_label} using the rerouting vector $\vg = \lb$.}
\label{fig:ce-improved}
\vspace{-1em}
\end{figure}

\begin{example}
\label{ex:improve-cegis-2}
Reconsider $\fml$ in Fig.~\ref{fig:my_label} with $\phi \equiv \safety{0.3}{\{t\}}$. 
Using the method outlined below we get: $\lb = [\sinit \mapsto 0.2, s_1 \mapsto 0.6, s_2 \mapsto 0.2, t \mapsto 1, f \mapsto 0]$. 
In absence of any bounds, the CE is $\{s_0, s_1, t\}$.  
Consider the gradual rerouting approach:
We set $\vg = \lb$, $C^{(0)} = \emptyset$ and have $D^{(0)} \coloneqq \subchain{\fml_{r_0}}{C^{(0)}}[\vg]$, see Fig.~\ref{fig:ce-improved}(a). 
Verifying this MC against $\phi' = \reachability{\leq}{0.3}{T\cup\{\sTop\}}$ yields $\prob[D^{(0)},\sinit \models \F{T\cup\{\sTop\}}] = \vg(\sinit) = 0.2 \leq 0.3$, i.e., the set~$C^{(0)}$ is not a CE. 
We now expand the initial state, i.e., $C^{(1)} = \{\sinit\}$ and let $D^{(1)} \coloneqq \subchain{\fml_{r_0}}{C^{(1)}}[\vg]$, see Fig.~\ref{fig:ce-improved}(b). 
Verifying $D^{(1)}$ yields $\prob[D^{(1)},\sinit \models \F{T\cup\{\sTop\}}] = 1 \cdot \vg(s_1) = 0.6 > 0.3$. 
Thus, the set $C^{(1)}$ is critical and the corresponding conflict is $K_{C^{(1)}} = \mathrm{supp}(\sinit) = \{X\}$. 
This is smaller than the naively computed conflict $\{X,Y\}$.
\end{example}

\setlength{\textfloatsep}{10pt}
\subsubsection{Greedy state expansion strategy}
Recall from Fig.~\ref{fig:expansion} that for an MC $\fmlr$ with $\fmlr \not\models \phi$, multiple CEs may exist inducing different conflicts. 
An efficient expansion strategy should yield a CE that induces a small amount of relevant parameters (to prune more family members) and this CE is preferably obtained by a small number of model-checking queries.
The method presented in Alg.~\ref{alg:ce} meets these criteria. 
\begin{algorithm}[t]
\SetKwInOut{Input}{Input}
\SetKwInOut{Output}{Output}
\Input{An MC $\fmlr$ a property $\phi \equiv \reachability{\bowtie}{\lambda}{T}$~s.t.~$\fmlr \not \models \phi$, a~rerouting vector $\vg$.}
\Output{A conflict $K$ for $\fmlr$ and $\phi$.}

\BlankLine
$i \gets 0$, $K^{(i)} \gets \emptyset$ \\
\While{true} {
    $C^{(i)}, H^{(i)} \gets \mathrm{reachableViaHoles}(\fmlr,K^{(i)})$ \label{alg:ce:reachable}\\
    $D^{(i)} \gets \subchain{\fmlr}{C^{(i)}}[\vg]$ \\
    \lIf{$\prob[D^{(i)} \models \F{T\cup\{\sTop\}}] \not \bowtie \lambda $} {\Return{$K^{(i)}$}}
    $\overline{s} \gets \mathrm{chooseToExpand}(H^{(i)},K^{(i)})$ \\
    $K^{(i+1)} = K^{(i)} \cup \supp(\fpm(\overline{s}))$ \\
    $i \gets i+1$\\
}

\caption{Counterexample construction based on rerouting.}
\label{alg:ce}
\end{algorithm}
The algorithm expands multiple states between subsequent model checks, while expanding only states that are associated with parameters that are relevant. 
In particular, in each iteration, we keep track of the set $K^{(i)}$ of relevant parameters optimistically starting with $K^{(0)} = \emptyset$.
We compute (see line~\ref{alg:ce:reachable}) the set $C^{(i)}$ of states that are reachable from the initial state via states which are associated only with relevant parameters in $K^{(i)}$, i.e., via states for which $\supp(\fpm(s)) \subseteq K^{(i)}$. 
Here, $H^{(i)}$ represents a state exploration `horizon': the set of states reachable from~$C^{(i)}$ but containing some (still) irrelevant parameters. 
We then construct the corresponding rerouting $\subchain{D}{C^{(i)}}[\vg]$ and check whether it is a CE. 
Otherwise, we greedily choose a state $\overline{s}$ from the horizon $H^{(i)}$ containing the least number of irrelevant parameters and add these parameters to our conflict~(see~line~7). 
The resulting conflict may not be minimal, but is computed fast. 
Our algorithm applies to probabilistic liveness properties\footnote{Some care is required regarding loops, see~\cite{cegis}.} too using $\vg = \ub$.

\subsubsection{Computing bounds} 
We compute $\lb$ and $\ub$ using an abstraction~\cite{cegar}.  
The method considers some set $\rlz$ of realizations and computes the corresponding \emph{quotient Markov decision process (MDP)} that over-approximates the behavior of all MCs in the family $\rlz$. 
Model checking this MDP yields an upper and a lower bound of the induced probabilities for all states over all realizations in $\rlz$. 
That is,
$\textsl{Bound}(\fml,\rlz)$ computes $\lb \in \mathbb{R}^S$ and $\ub \in \mathbb{R}^S$ such that for each $\sins$:
\[ 
\lb(s) \ \leq \ 
\min_{r \in \rlz} \prob[\fmlr, s \models \F{T}] \ \leq \ 
\max_{r \in \rlz} \prob[\fmlr, s \models \F{T}] \ \leq \ 
\ub(s).  
\]  
To allow for refinement, two properties are crucial (with point-wise inequalities):
\[\mbox{1. } \lb \leq  \textsl{lb}^{\rlz'} \wedge  \ub \geq  \textsl{ub}^{\rlz'}  \mbox{ for } \rlz' \subseteq \rlz \quad \mbox{and}\quad \mbox{2. } \textsl{lb}^{\{ r \}}  =  \textsl{ub}^{\{ r \}} \mbox{ for } r \in \rlz. \]
In~\cite{cegar}, the abstraction and refinement together define an abstraction-refinement loop (AR) that addresses the feasibility problem. 
In the worst case, this loop analyses $2\cdot|\rlz|$ quotient MDPs, which (as of now)  may be arbitrarily larger than the number of family members they represent. 

\section{Hybrid Dual-Oracle Synthesis}
\label{sec:adaptive}
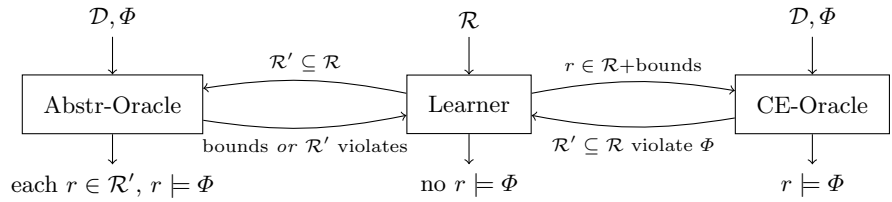
\begin{figure}[t]
    \centering
     \begin{tikzpicture}
    \node[rectangle, draw, inner sep=8pt] (learner) {Learner};
    \node[rectangle, draw, inner sep=8pt,right=2.7cm of learner] (oracle) {CE-Oracle};
       \node[rectangle, draw, inner sep=8pt,left=2.7cm of learner] (abst) {Abstr-Oracle};
    \node[above=0.5cm of learner] (rlz) {$\rlz$};
    \draw[->] (rlz) -- (learner);
    \node[above=0.5cm of oracle] (phi) {$\mathcal{D},\Phi$};
    \draw[->] (phi) -- (oracle);
    \node[above=0.5cm of abst] (phiab) {$\mathcal{D},\Phi$};
    \draw[->] (phiab) -- (abst);
    \draw[->] (learner) edge[bend left=10] node[above] {\scriptsize{$r \in \rlz$}+bounds} (oracle);
    \draw[->] (oracle) edge[bend left=10] node[below] {\scriptsize{$\rlz' \subseteq \rlz$ violate $\Phi$}} (learner);
      \draw[->] (learner) edge[bend right=10] node[above] {\scriptsize{$\rlz' \subseteq \rlz$}} (abst);
    \draw[->] (abst) edge[bend right=10] node[below,align=center] {\scriptsize{bounds \emph{or} $\rlz'$ violates}} (learner);
    
    \node[below=0.4cm of oracle] (sat) {$r \models \Phi$};
    \draw[->] (oracle) -- (sat);
    \node[below=0.4cm of abst] (allsat) {each $r \in \rlz'$, $r \models \Phi$};
    \draw[->] (abst) -- (allsat);
    \node[below=0.4cm of learner] (unsat) {no $r \models \Phi$};
    \draw[->] (learner) -- (unsat);
    \end{tikzpicture}
    \caption{Conceptual hybrid (dual-oracle) synthesis}
    \label{fig:adaptivesynt}.
    \vspace{-0.5em}
\end{figure}

We introduce an extended synthesis loop in which the abstraction-based reasoning is used to prune the family $\rlz$, and to accelerate the CE-based oracle from Sect.~\ref{sec:main}. 
The intuitive idea is outlined in Fig.~\ref{fig:adaptivesynt}.
Note that if the CE-based oracle is not exploited, we emulate AR (explained in computing bounds above), whereas if the abstraction oracle is not used, we emulate CEGIS (with the novel oracle). 

Let us motivate combining these oracles in a flexible way. 
The naive version outlined in the previous section assumed a single abstraction step, and invokes CEGIS with the bounds obtained from that step. 
Evidently, the better (tighter) the bounds~$\vg$, the better the CEs. 
However, the abstraction-based bounds for $\rlz$ may be very loose. 
These bounds can be improved by splitting the set $\rlz$ and using the bounds on the two sub-families. 
The idea is to run a limited number of AR steps and then invoke CEGIS. 
Our experiments reveal that it can be crucial to be adaptive, i.e., the integrated method must be able to detect at run time when to switch. 

\begin{algorithm}[t]
\SetKwInOut{Input}{Input}
\SetKwInOut{Output}{Output}
\SetKwFunction{FMain}{adaptiveSynthesis}
\SetKwProg{Fn}{Function}{:}{}
\Input{A family $\fml$, a~reachability property $\phi$.}
\Output{Either a member $r$ in $\fml$ with $r \models \phi$, or no such $r$ exists in $\fml$}
{
\BlankLine
$\overline{\rlz} \gets \{\rlzf\}$ \tcp*{each analysed (sub-)family also holds bounds}
$\delta_{CEGIS} \gets 1$ \tcp*{time allocation factor for CEGIS}
\While{true} {
    result,$\overline{\rlz}',\successRate{AR},t_{AR} \gets $AR.run$(\overline{\rlz},\phi)$ \\
    \lIf{$result.\mathrm{decided}()$} {\Return{$result$}}
    CEGIS.setTimeout($t_{AR} \cdot \delta_{CEGIS}$) \\
    $result,\successRate{CEGIS},\overline{\rlz}'' \gets \mathrm{CEGIS.run}(\overline{\rlz}',\phi)$ \\
    \lIf{$result.\mathrm{decided}()$} {\Return{$result$}}
    $\delta_{CEGIS} \gets \successRate{CEGIS} / \successRate{AR}$ \\
    $\overline{\rlz} \gets \overline{\rlz}''$ \\
}
}
\caption{Hybrid (dual-oracle) synthesis.}
\label{alg:hybrid}
\end{algorithm}
The proposed hybrid method switches between AR and CEGIS, where we allow for refining during the AR phase and use the obtained refined bounds during CEGIS. 
Additionally, we estimate the efficiency~$\sigma$ (e.g., the number of pruned MCs per time unit) of the two methods and allocate more time $t$ to the method with superior performance. 
That is, if we detect that CEGIS prunes sub-families twice as fast as AR, we double the time in the next round for CEGIS. 
The resulting algorithm is summarized in Alg.~\ref{alg:hybrid}. Recall that AR (at line 5) takes one family from $\overline{\rlz}$, either solves it or splits it and returns the set of undecided families $\overline{\rlz}'$. 
In contrast, CEGIS processes multiple families from $\overline{\rlz}'$ until the timeout and then returns the set of undecided families $\overline{\rlz}''$. 
This workflow is motivated by the fact that one iteration of AR (i.e., the involved MDP model-checking) is typically significantly slower that one CEGIS iteration.

\begin{remark}
\label{remark:gen}
Although the developed framework for integrated synthesis has been discussed in the context of feasibility with respect to~a~single property $\phi$, it can be easily generalized to handle \emph{multiple}-property specifications as well as to treat \emph{optimal} synthesis. 
Regarding multiple properties, the idea remains the same: Analyzing the quotient MDP with respect to multiple properties yields multiple probability bounds. 
After initiating a CEGIS-loop and obtaining an unsatisfiable realization, we can construct a separate conflict for each unsatisfied property, while using the corresponding probability bound to enhance the CE generation process.
Optimal synthesis is handled similarly to feasibility, but, after obtaining a satisfiable solution, we update the optimizing property to exclude this solution: e.g., for maximal synthesis this translates to increasing the threshold of the maximizing property. Having exhausted the search space of family members, the last obtained solution is declared to be the optimal one.
\end{remark}

\section{Experimental evaluation}
\label{sec:experiments}

\paragraph{Implementation.}
We implemented the hybrid oracle
on top of the probabilistic model checker Storm~\cite{STORM}. 
While the high-performance parts were implemented in C++, we used a python API to flexibly construct the overall synthesis loop. 
For SMT solving, we used Z3~\cite{z3}. 
The~tool chain takes a~PRISM~\cite{KNP11} or JANI~\cite{jani} sketch and a set of temporal properties, and returns a~satisfying realization, if such exists, or outputs that such realization does not exist. The implementation in the form of an artefact is available at \url{https://zenodo.org/record/4422543}.

\paragraph{Set-up.}
We compare the adaptive oracle-guided synthesis with two state-of-the-art synthesis methods: program-level CEGIS~\cite{cegis} using a MaxSat CE generation~\cite{DBLP:conf/atva/DehnertJWAK14,DBLP:journals/corr/abs-1305-5055} and AR~\cite{cegar}. These use the same architecture and data structures from Storm. 
All experiments are run on an~Ubuntu~19.04 machine with Intel i5-8300H (4 cores at 2.3 GHz) and using up to 8 GB RAM, with all the~algorithms being executed on a single~thread. 
The~benchmarks consists of five different models, see Table~\ref{tab:benchmark}, from various domains that were used in~\cite{cegis,cegar}.  
As opposed to the benchmark considered in~\cite{cegis,cegar}, we use larger variants of \emph{Grid} and \emph{Herman} to better demonstrate differences in the performance of individual methods.

To investigate the scalability of the methods, we consider a new variant of the \emph{Herman} model, that allows us to scale the number of randomization strategies and thus the family size.
In particular, we will compare performance on two instances of different sizes: \emph{small \hermantwo} (5k members) and \emph{large \hermantwo} (3.1M members, other statistics are reported in Table~\ref{tab:benchmark}).

\begin{table}[t]
\begin{center}
\begin{tabular} {l|c|c|c|c|}
model & $|K|$ & $|\rlzf|$ & MDP size & avg.~MC size \\ \hline
\emph{Grid} & 8 & 65k & 11.5k & 1.2k \\
\emph{Maze} & 20 & 1M & 9k & 5.4k \\
\emph{DPM} & 16 & 43M & 9.5k & 2.2k 
\end{tabular} \hspace{-0.4em}
\begin{tabular} {|l|c|c|c|c}
model & $|K|$ & $|\rlzf|$ & MDP size & avg.~MC size \\ \hline
\emph{Pole} & 17 & 1.3M & 6.6k & 5.6k \\
\emph{Herman} & 8 & 0.5k & 48k & 5.2k \\
\emph{Herman$^*$} & 7 & 3.1M & 6k & 1k
\end{tabular}
\vspace{0.5em}
\caption{Summary of the benchmarks and their statistics}
\label{tab:benchmark}
\end{center}
\vspace{-2em}
\end{table}

To reason about the pruning efficiency of different synthesis methods, we want to avoid feasible synthesis problems, where the order of family exploration can lead to inconsistent performance. Instead, we will primarily focus on non-feasible problems, where all realizations need to be explored in order to prove unsatisfiability.
The experimental evaluation is presented in three parts.
(1) We evaluate the novel CE construction method and compare it with the MaxSat-based oracle from~\cite{cegis}. 
(2) We compare the hybrid synthesis loop with the two baselines AR and CEGIS. 
(3) We consider novel hard synthesis instances (multi-property synthesis, finding optimal programs) on instances of the model \emph{\hermantwo}.

\subsubsection{Comparing CE construction methods}
We consider \emph{the quality of the CEs} and \emph{their generation time}.
In particular, we want to investigate (1) whether using CEs-modulo-families yields better CEes, (2) how the quality of CEs from the smart oracle compares to the MaxSat-based oracle, and how
their time consumption compares.
As a measure of quality of a CE, the average number of its relevant parameters w.r.t.~the total number of its parameters is taken.
That is, smaller ratios imply better CEs.
To measure the influence of using CEs-modulo-families, two types of bounds are used: (i)~trivial bounds (i.e., $\vg=\vz$ for safety and $\vg = \vo$ for liveness properties), and (ii)~non-trivial bounds corresponding to the entire family $\rlzf$ representing the most conservative estimate. 
The results are reported in (the left part of) Table~\ref{tab:ce}.
In the next subsection, we investigate this same benchmark from the point of view of the performance of the synthesis methods, which also shows the immediate effect of the new CE generation strategy.

\begin{table}[t]
\begin{center}
\begin{tabular} {lr|c|c|c|c|c|c|c|c|c}
\multicolumn{2}{c|}{\multirow{3}{*}{model}} & \multicolumn{3}{c|}{CE quality} & \multicolumn{6}{c}{performance} \\
& & \multirow{2}{*}{MaxSat~\cite{DBLP:conf/atva/DehnertJWAK14}} & \multicolumn{2}{c|}{state expansion (\textcolor{red}{new})} & \multicolumn{2}{c|}{CEGIS~\cite{cegis}} & \multicolumn{2}{c|}{AR~\cite{cegar}} & \multicolumn{2}{c}{Hybrid (\textcolor{red}{new})} \\
& & & trivial & non-trivial & iters & time & iters & time & iters & time \\
\hline
\multirow{2}{*}{\emph{Grid}} & & 0.59 (0.025) & 0.50 (0.001) & 0.50 & 613 & 30 & 5325& 486 & (285, 11) & \textbf{6} \\
& $*$ & 0.74 (0.026) & 0.65 (0.001) & 0.65 & 1801 & 93 & 6139 & 540 & (2100, 127) & \textbf{33} \\ \hline
\multirow{2}{*}{\emph{Maze}} & &  0.21 (0.247) & 0.55 (0.009) &  0.38  & 290 & 5449 & 49 & 17 & (105, 13) & \textbf{7} \\
& $*$ &   0.24 (2.595) & 0.63 (0.012) & 0.46  & 301 & 6069 & 63 & 26 & (146, 17) & \textbf{9} \\ \hline
\multirow{2}{*}{\emph{DPM}} & & 0.32 (0.447) & 0.61 (0.007) & 0.53  & 2906 & 2488 & 299 & 25 & (631, 143) & \textbf{23} \\
& $*$ & 0.33 (0.525) & 0.49 (0.006) & 0.42  & 3172 & 2782 & 1215 & 81 & (2374, 545) & \textbf{76} \\ \hline
\multirow{2}{*}{\emph{Pole}} & &  - & 0.87 (0.062) & 0.16  & - & - & 309 & 12 & (3, 5) & \textbf{1} \\
& $*$& - & 0.54 (0.041) & 0.29  & - & - & 615 & 23 & (80, 61) & \textbf{6} \\ \hline
\multirow{2}{*}{\emph{Herman}} & & - & 0.91 (0.011) & 0.50 & - & - & 171 & 86 & (24, 1) & \textbf{9} \\
& $*$ & - & 0.88 (0.016) & 0.87  & - & - & 643 & 269 & (485, 13) & \textbf{29} \\
\end{tabular}
\vspace{0.6em}
\caption{
CE quality for different methods and performance of three synthesis methods. For each model/property, we report results for two different thresholds where the symbol `$*$' marks the one closer to the feasibility threshold, representing the more difficult synthesis problem. Symbol `-' marks a two-hour timeout. \textbf{CE~quality}: The presented numbers give the CE quality (i.e., the smaller, the better). The numbers in parentheses represent the average run-time of constructing one CE in seconds
(run-times for constructing CE using non-trivial bounds are similar as for trivial ones and are thus not reported).
\textbf{Performance}: for each method, we report the  number of iterations (for the hybrid method, the reported values are iterations of the CEGIS and AR oracle, respectively) and the run-time in seconds.
}
\label{tab:ce}
\end{center}
\vspace{-1em}
\end{table}

The first observation is that using non-trivial bounds (as opposed to trivial ones) for the state expansion approach can drastically decrease the conflict size.
It turns out that the CEs obtained using the greedy approach are mostly larger than those obtained with the MaxSat method.
However (see \emph{Grid}), even for trivial bounds, we may obtain smaller CEs than for MaxSat: computing a minimal-command CE does not necessarily induce an optimal conflict. 
On the other hand, comparing the run-times in the parentheses, one can see that computing CEs via the greedy state expansion is orders of magnitude faster than computing command-optimal ones using MaxSat.
It is good to realize that the greedy method makes at most $|K|$ model-checking queries to compute CEs, while the MaxSat method may make exponentially many such queries.
Overall, the greedy method using the non-trivial bounds is able to obtain CEs of comparable quality as the MaxSat method, while being orders of magnitude faster.

\subsubsection{Performance comparison with AR/CEGIS}
We compare the  hybrid synthesis loop from Sect.~\ref{sec:adaptive} with two state-of-the-art baselines: CEGIS and AR.
The results are displayed in (the right half of) Table~\ref{tab:ce}.
\emph{In all 10 cases, the hybrid method outperforms the baselines. It is up to an order of magnitude faster}.

Let us discuss the performance of the hybrid method.
We classify benchmarks along two dimensions: (1) the performance of CEGIS and (2) the performance of AR. 
Based on the empirical performance, we classify (\emph{Grid}) as good-for-CEGIS (and not for AR), \emph{Maze}, \emph{Pole} and \emph{DPM} as good-for-AR (and not for CEGIS), and \emph{Herman} as hard (for both).
Roughly, 
AR works well when the quotient MDP does not blow up and its analysis is precise due to consistent schedulers, i.e., when the parameter dependencies are not crucial for a precise analysis.
CEGIS performs well when the CEs are small and fast to compute.
On the other hand, synthesis problems for which neither pure CEGIS nor pure AR are able to effectively reason about non-trivial subfamilies, inherently profit from a hybrid method.
The main point we want to discuss is \emph{how the hybrid method reinforces the strengths of both methods, rather than their weaknesses}.

In the hybrid method, there are two factors that determine the efficiency: (i)~\emph{how fast do we get bounds on the reachability probability that are tight enough to enable construction of good counterexamples?} and (ii) \emph{how good are the constructed counterexamples?} 
The former factor is attributed to the proposed adaptive scheme (see Alg.~\ref{alg:hybrid}), where the method will prefer AR-like analysis and continue refinement until the computed bounds allow construction of small counterexamples. 
The latter is reflected above. 
Let us now discuss how these two aspects are reflected in the  benchmarks.

In good-for-CEGIS benchmarks like \emph{Grid}, after analyzing a quotient MDP for the whole family, the hybrid method mostly profits from better CEs yielding better bounds, thus outperforming CEGIS. Indeed, the CEs are found so fast that the bottleneck is no longer their generation. 
This also explains why the speedup is not immediately translated to the speedup on the overall synthesis loop.
In the good-for-AR benchmark \emph{DPM}, the hybrid method provides only a minor improvement as it has to perform a large number of AR-iterations before the novel CE-based pruning can be effectively used. 
This can be considered as the worst-case scenario for the hybrid method.
On other good-for-AR benchmarks like \emph{Maze} and \emph{Pole}, the good performance on AR allows to quickly obtain tight bounds which can then be exploited by CEGIS.
Finally, in hard models like \emph{Herman}, abstraction-refinement is very expensive, but even the bounds from the first round yield bounds that, as opposed to the trivial bounds, now enable good CEs: CEGIS can keep using these bounds to quickly prune the state space.

\vspace{-1em}
\subsubsection{More complicated synthesis problems}
Our new approach can push the limits of synthesis benchmarks significantly. 
We illustrate this by considering a new variant of the \emph{Herman} model, \emph{\hermantwo}, and a property imposing an upper bound on the expected number of rounds until stabilization. 
We put this bound just below the optimal (i.e., the minimal) value, yielding a hard non-feasible problem.
The synthesis results are summarized in Table~\ref{tab:advanced}.
As CEGIS performs poorly on \emph{Herman}, it is excluded here. 

First, we investigate on \emph{small \hermantwo} how the methods can handle the synthesis for multi-property specifications. We add one feasible property to the (still non-feasible) specification (row `two properties').
While including more properties typically slows down the AR computation, the performance of the hybrid method is not affected as the corresponding overhead is mitigated by additional pruning opportunities. 
Second, we consider optimal synthesis for the property as used in the feasibility synthesis. 
The hybrid method requires only a minor overhead to find an optimal solution compared to checking feasibility.
This overhead is significantly larger for AR. 

Next, we consider \emph{larger \hermantwo} model having significantly more randomization strategies (3.1M members) that include solutions leading to a considerably faster stabilization.
This model is out of reach for existing synthesis approaches: one-by-one enumeration takes more than 27 hours and the AR performs even worse---solving the feasibility and optimality problems requires 47 and 55 hours, respectively. 
On the other hand, the proposed hybrid method is able to solve these problems within minutes.
Finally, we consider a relaxed variant of optimal synthesis (\mbox{$5\%$-optimality}) guaranteeing that the found solution is up to $5\%$ worse than the optimal. 
Relaxing the optimally criterion speeds up the hybrid synthesis method by about a factor three.

These experiments clearly demonstrate that scaling up the synthesis problem several orders of magnitude renders existing synthesis methods infeasible: they need tens of hours to solve the synthesis problems.
Meanwhile, the hybrid method tackles these difficult synthesis problems without significant~penalty and is capable of producing a solution within minutes.

\begin{table}[t]
\begin{tabular} {|l|c|c|c|c|}
synthesis & \multicolumn{2}{c|}{AR} & \multicolumn{2}{c|}{Hybrid} \\
problem &  iters & time & iters & time  \\
\hline
feasibility  & 81 & 30s & (274, 1) & \textbf{7s} \\
two properties & 97 & 38s & (274, 1) & \textbf{8s} \\
optimality & 531 & 150s & (571, 7) & \textbf{12s} \\

\end{tabular} \hfill
$\quad$
\begin{tabular} {|l|c|c|c|c}
synthesis & \multicolumn{2}{c|}{AR} & \multicolumn{2}{c}{Hybrid} \\
problem &  iters & time & iters & time  \\
\hline

feasibility  & 69k & 47h & (14280, 2) & \textbf{13.4m} \\
optimality & 83k & 55h & (16197, 3) & \textbf{16.8m} \\
$5\%$-optimality  & 60k & 42h & (6421, 7) &\textbf{5.1m} \\

\end{tabular} \hfill
\vspace{0.5em}
\caption{The impact of scaling the family size (of the \emph{\hermantwo} model) and handling more complex synthesis problems.
The left part shows the results for the smaller variant (5k members), the right part for the larger one (3.1M members). }
\label{tab:advanced}
\vspace{-0.5em}
\end{table}

\section{Conclusion}
We present a novel method for the automated synthesis of probabilistic programs. Pairing the counterexample-guided inductive synthesis with the deductive oracle using an MDP abstraction, we develop a synthesis technique 
enabling faster construction of smaller counterexamples. Evaluating the method on  case studies from different domains, we demonstrate that 
the novel CE construction and the adaptive strategy
lead to a significant acceleration of the synthesis process. The proposed method is able to reduce the run-time for challenging problems from days to minutes. In our future work, we plan to investigate counterexamples on the quotient MDPs and 
improve the abstraction refinement strategy. 

\clearpage\pagebreak
\bibliographystyle{splncs04}
\bibliography{bibliography}

\begin{thebibliography}{10}
\providecommand{\url}[1]{\texttt{#1}}
\providecommand{\urlprefix}{URL }
\providecommand{\doi}[1]{https://doi.org/#1}

\bibitem{DBLP:conf/sfm/AbrahamBDJKW14}
{\'{A}}brah{\'{a}}m, E., Becker, B., Dehnert, C., Jansen, N., Katoen, J.P.,
  Wimmer, R.: Counterexample generation for discrete-time {M}arkov models: An
  introductory survey. In: {SFM}. LNCS, vol.~8483, pp. 65--121. Springer (2014)

\bibitem{sygus}
Alur, R., Bod{\'{\i}}k, R., Dallal, E., Fisman, D., Garg, P., Juniwal, G.,
  Kress{-}Gazit, H., Madhusudan, P., Martin, M.M.K., Raghothaman, M., Saha, S.,
  Seshia, S.A., Singh, R., Solar{-}Lezama, A., Torlak, E., Udupa, A.:
  Syntax-guided synthesis. In: Dependable Software Systems Engineering, {NATO}
  Science for Peace and Security Series, vol.~40, pp. 1--25. {IOS} Press (2015)

\bibitem{Baier2018}
Baier, C., de~Alfaro, L., Forejt, V., Kwiatkowska, M.: Model checking
  probabilistic systems. In: Handbook of Model Checking, pp. 963--999. Springer
  (2018)

\bibitem{DBLP:journals/iandc/BaierHHJKK20}
Baier, C., Hensel, C., Hutschenreiter, L., Junges, S., Katoen, J., Klein, J.:
  Parametric markov chains: {PCTL} complexity and fraction-free gaussian
  elimination. Inf. Comput.  \textbf{272},  104504 (2020)

\bibitem{model-repair-1}
Bartocci, E., Grosu, R., Katsaros, P., Ramakrishnan, C.R., Smolka, S.A.: Model
  repair for probabilistic systems. In: {TACAS'11}. LNCS, vol.~6605, pp.
  326--340 (2011)

\bibitem{jani}
Bornholt, J., Torlak, E., Grossman, D., Ceze, L.: Optimizing synthesis with
  metasketches. In: POPL'16. p. 775–788. Association for Computing Machinery
  (2016)

\bibitem{CALINESCU2018140}
Calinescu, R., {\v{C}}e\v{s}ka, M., Gerasimou, S., Kwiatkowska, M., Paoletti,
  N.: Efficient synthesis of robust models for stochastic systems. J.\ of
  Systems and Softw.  \textbf{143},  140--158 (2018)

\bibitem{acta}
{\v{C}}e{\v{s}}ka, M., Dannenberg, F., Paoletti, N., Kwiatkowska, M., Brim, L.:
  Precise parameter synthesis for stochastic biochemical systems. Acta Inf.
  \textbf{54}(6),  589--623 (2017)

\bibitem{cegis}
{\v{C}}e{\v{s}}ka, M., Hensel, C., Junges, S., Katoen, J.P.:
  Counterexample-driven synthesis for probabilistic program sketches. In: {FM}.
  LNCS, vol. 11800, pp. 101--120. Springer (2019)

\bibitem{cegar}
{\v{C}}e{\v{s}}ka, M., Jansen, N., Junges, S., Katoen, J.P.: Shepherding hordes
  of {M}arkov chains. In: {TACAS} {(2)}. LNCS, vol. 11428, pp. 172--190.
  Springer (2019)

\bibitem{DBLP:journals/iandc/Chatzieleftheriou18}
Chatzieleftheriou, G., Katsaros, P.: Abstract model repair for probabilistic
  systems. Inf. Comput.  \textbf{259}(1),  142--160 (2018)

\bibitem{10.1007/978-3-030-30806-3_7}
Chonev, V.: Reachability in augmented interval {M}arkov chains. In: RP'2019.
  LNCS, vol. 11674, pp. 79--92. Springer (2019)

\bibitem{DBLP:journals/fac/ChrszonDKB18}
Chrszon, P., Dubslaff, C., Kl{\"{u}}ppelholz, S., Baier, C.: Pro{F}eat:
  feature-oriented engineering for family-based probabilistic model checking.
  Formal Asp. Comput.  \textbf{30}(1),  45--75 (2018)

\bibitem{onebyone}
Classen, A., Cordy, M., Heymans, P., Legay, A., Schobbens, P.Y.: Model checking
  software product lines with {SNIP}. Int.\ J.\ on Softw.\ Tools for Technol.
  Transf.  \textbf{14},  589--612 (2012)

\bibitem{DBLP:conf/ictac/Daws04}
Daws, C.: Symbolic and parametric model checking of discrete-time {M}arkov
  chains. In: {ICTAC}. LNCS, vol.~3407, pp. 280--294. Springer (2004)

\bibitem{DBLP:conf/atva/DehnertJWAK14}
Dehnert, C., Jansen, N., Wimmer, R., {\'{A}}brah{\'{a}}m, E., Katoen, J.P.:
  Fast debugging of {PRISM} models. In: {ATVA}. LNCS, vol.~8837, pp. 146--162.
  Springer (2014)

\bibitem{dehnert2015prophesy}
Dehnert, C., Junges, S., Jansen, N., Corzilius, F., Volk, M., Bruintjes, H.,
  Katoen, J.P., {\'A}brah{\'a}m, E.: {PROPhESY: A PRObabilistic ParamEter
  SYNnthesis Tool}. In: CAV'15. LNCS, vol.~9206, pp. 214--231. Springer (2015)

\bibitem{STORM}
Dehnert, C., Junges, S., Katoen, J.P., Volk, M.: A {Storm} is coming: A modern
  probabilistic model checker. In: CAV. LNCS, vol. 10427, pp. 592--600.
  Springer (2017)

\bibitem{DBLP:conf/tacas/FunkeJB20}
Funke, F., Jantsch, S., Baier, C.: Farkas certificates and minimal witnesses
  for probabilistic reachability constraints. In: {TACAS} {(1)}. LNCS, vol.
  12078, pp. 324--345. Springer (2020)

\bibitem{DBLP:journals/ase/GerasimouCT18}
Gerasimou, S., Calinescu, R., Tamburrelli, G.: Synthesis of probabilistic
  models for quality-of-service software engineering. Autom. Softw. Eng.
  \textbf{25}(4),  785--831 (2018)

\bibitem{DBLP:journals/infsof/GhezziS13}
Ghezzi, C., Sharifloo, A.M.: Model-based verification of quantitative
  non-functional properties for software product lines. Inf. {\&} Softw.
  Technol.  \textbf{55}(3),  508--524 (2013)

\bibitem{hahn2011probabilistic}
Hahn, E.M., Hermanns, H., Zhang, L.: Probabilistic reachability for parametric
  {M}arkov models. Int.\ J.\ on Softw.\ Tools for Technol. Transf.
  \textbf{13}(1),  3--19 (2011)

\bibitem{Harman2012}
Harman, M., Mansouri, S.A., Zhang, Y.: Search-based software engineering:
  Trends, techniques and applications. ACM Comp. Surveys  \textbf{45}(1),
  11:1--11:61 (2012)

\bibitem{herman-1}
Herman, T.: Probabilistic self-stabilization. Inf. Process. Lett.
  \textbf{35}(2),  63–67 (1990)

\bibitem{10.1145/1806799.1806833}
Jha, S., Gulwani, S., Seshia, S.A., Tiwari, A.: Oracle-guided component-based
  program synthesis. In: ICSE. p. 215–224. ACM (2010)

\bibitem{herman-2}
Kwiatkowska, M., Norman, G., Parker, D.: Probabilistic verification of
  {H}erman’s self-stabilisation algorithm. Formal Aspects of Computing
  \textbf{24}(4),  661--670 (2012)

\bibitem{KNP11}
Kwiatkowska, M., Norman, G., Parker, D.: {PRISM} 4.0: Verification of
  probabilistic real-time systems. In: CAV. LNCS, vol.~6806, pp. 585--591.
  Springer (2011)

\bibitem{lanna2018feature}
Lanna, A., Castro, T., Alves, V., Rodrigues, G., Schobbens, P.Y., Apel, S.:
  Feature-family-based reliability analysis of software product lines. Inf.\
  and Softw.\ Technol.  \textbf{94},  59--81 (2018)

\bibitem{z3}
Lindemann, C.: Performance modelling with deterministic and stochastic {P}etri
  nets. SIGMETRICS Perform. Eval. Rev.  \textbf{26}(2), ~3 (1998)

\bibitem{DBLP:conf/aaai/MadaniHC99}
Madani, O., Hanks, S., Condon, A.: On the undecidability of probabilistic
  planning and infinite-horizon partially observable {M}arkov decision
  problems. In: {AAAI/IAAI}. pp. 541--548. {AAAI} Press / The {MIT} Press
  (1999)

\bibitem{Martens:2010:AIS:1712605.1712624}
Martens, A., Koziolek, H., Becker, S., Reussner, R.: Automatically improve
  software architecture models for performance, reliability, and cost using
  evolutionary algorithms. In: WOSP/SIPEW. pp. 105--116. ACM (2010)

\bibitem{NoriPLDI2015}
Nori, A.V., Ozair, S., Rajamani, S.K., Vijaykeerthy, D.: Efficient synthesis of
  probabilistic programs. In: PLDI'14. pp. 208--217. ACM (2015)

\bibitem{DBLP:series/sbis/OliehoekA16}
Oliehoek, F.A., Amato, C.: A Concise Introduction to Decentralized POMDPs.
  Springer Briefs in Intelligent Systems, Springer (2016)

\bibitem{pathak-et-al-nfm-2015}
Pathak, S., {\'{A}}brah{\'{a}}m, E., Jansen, N., Tacchella, A., Katoen, J.P.: A
  greedy approach for the efficient repair of stochastic models. In: {NFM'15}.
  LNCS, vol.~9058, pp. 295--309. Springer (2015)

\bibitem{Put94}
Puterman, M.L.: Markov Decision Processes: Discrete Stochastic Dynamic
  Programming. Wiley Series in Probability and Statistics, Wiley (1994)

\bibitem{Quatmann2016}
Quatmann, T., Dehnert, C., Jansen, N., Junges, S., Katoen, J.P.: Parameter
  synthesis for {M}arkov models: Faster than ever. In: {ATVA'16}. LNCS,
  vol.~9938, pp. 50--67 (2016)

\bibitem{ce-rewards}
Quatmann, T., Jansen, N., Dehnert, C., Wimmer, R., {\'A}brah{\'a}m, E., Katoen,
  J.P., Becker, B.: Counterexamples for expected rewards. In: FM. pp. 435--452.
  Springer (2015)

\bibitem{saad2019bayesian}
Saad, F.A., Cusumano-Towner, M.F., Schaechtle, U., Rinard, M.C., Mansinghka,
  V.K.: Bayesian synthesis of probabilistic programs for automatic data
  modeling. Proceedings of the ACM on Programming Languages  \textbf{3}(POPL),
  1--32 (2019)

\bibitem{Solar-LezamaPLDI2005}
Solar-Lezama, A., Rabbah, R., Bod\'{\i}k, R., Ebcio\u{g}lu, K.: Programming by
  sketching for bit-streaming programs. In: PLDI'05. pp. 281--294. ACM (2005)

\bibitem{DBLP:conf/fm/VandinBLL18}
Vandin, A., ter Beek, M.H., Legay, A., Lluch{-}Lafuente, A.: Qflan: {A} tool
  for the quantitative analysis of highly reconfigurable systems. In: {FM}.
  LNCS, vol. 10951, pp. 329--337. Springer (2018)

\bibitem{DBLP:journals/corr/abs-1305-5055}
Wimmer, R., Jansen, N., Vorpahl, A., {\'{A}}brah{\'{a}}m, E., Katoen, J.P.,
  Becker, B.: High-level counterexamples for probabilistic automata. Logical
  Methods in Computer Science  \textbf{11}(1) (2015)

\end{thebibliography}
%


\vfill

{\small\medskip\noindent{\bf Open Access} This chapter is licensed under the terms of the Creative Commons\break Attribution 4.0 International License (\url{http://creativecommons.org/licenses/by/4.0/}), which permits use, sharing, adaptation, distribution and reproduction in any medium or format, as long as you give appropriate credit to the original author(s) and the source, provide a link to the Creative Commons license and indicate if changes were made.}

{\small \spaceskip .28em plus .1em minus .1em The images or other third party material in this chapter are included in the chapter's Creative Commons license, unless indicated otherwise in a credit line to the material.~If material is not included in the chapter's Creative Commons license and your intended\break use is not permitted by statutory regulation or exceeds the permitted use, you will need to obtain permission directly from the copyright holder.}

\medskip\noindent\includegraphics{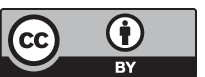}

\end{document}